\documentclass[aps,twocolumn,pra,superscriptaddress,letterpaper]{revtex4-1}
\usepackage{amsmath,amsfonts,amssymb,amsthm,mathalfa}
\usepackage{mathtools}
\usepackage[pdftex]{graphicx}
\usepackage[usenames, dvipsnames, svgnames, table]{xcolor}
\usepackage[colorlinks=true, citecolor=Blue, linkcolor=BrickRed, urlcolor=Brown]{hyperref}
\usepackage{txfonts}
\usepackage{mathrsfs}
\usepackage{bm}
\usepackage{multirow}
\usepackage{braket}
\usepackage{import}
\usepackage[caption=false]{subfig}
\usepackage[boxed,noline,noend]{algorithm2e}
\usepackage{array}
\usepackage{comment}
\usepackage{diagbox, eqparbox, hhline}

\usepackage{tikz}
\usetikzlibrary{quantikz}

\newtheorem{theorem}{Theorem}
\newtheorem{lemma}{Lemma}
\theoremstyle{definition}
\newtheorem{definition}{Definition}
\newcommand{\be}{\begin{equation}}
\newcommand{\ee}{\end{equation}}
\newcommand{\ben}{\begin{eqnarray}}
\newcommand{\een}{\end{eqnarray}}
\newcommand{\bes}{\begin{subequations}}
\newcommand{\ees}{\end{subequations}}
\newcommand{\bF}{\begin{figure}}
\newcommand{\eF}{\end{figure}}

\DeclareMathOperator{\tr}{Tr}

\SetKwInput{KwInput}{Input}
\SetKwInput{KwOutput}{Output}

\DeclareMathAlphabet{\pazocal}{OMS}{zplm}{m}{n}

\newcommand{\A}{\pazocal{A}}

\newcommand{\G}{\pazocal{G}}

\newcommand{\D}{\pazocal{D}}
\newcommand{\s}{\pazocal{S}}
\newcommand{\B}{\pazocal{B}}
\newcommand{\oo}{\pazocal{O}}

\newcommand{\U}{\pazocal{U}}

\newcommand{\w}{\pazocal{W}}

\newcommand{\FP}{\mathfrak{P}}
\newcommand{\FI}{\mathcal{I}}

\newcommand{\TR}{\mathcal{T}_R}
\newcommand{\TI}{\mathcal{T}_I}

\newcommand{\zt}{Z_{\mathrm{T}}}

\newcommand{\oTR}{\mathcal{T}_{R,1}}
\newcommand{\oTI}{\mathcal{T}_{I,1}}

\newcommand{\oT}{\mathcal{T}_{1}}

\newcommand{\me}{\epsilon_{\mathrm{m}}}
\newcommand{\abse}{\epsilon_{\mathrm{a}}}

\newcommand{\ase}{\epsilon_{\mathrm{aS}}}
\newcommand{\ate}{\epsilon_{\mathrm{aT}}}

\newcommand{\mse}{\epsilon_{\mathrm{mS}}}
\newcommand{\mte}{\epsilon_{\mathrm{mT}}}

\def\SN#1{ \left \vert \left \vert #1 \right \vert \right \vert}

\newcommand{\polylog}{\mathrm{polylog}}
\newcommand{\poly}{\mathrm{poly}}

\newcommand{\orcid}[1]{\href{https://orcid.org/#1}{\includegraphics[height = 2ex]{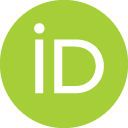}}}

\begin{document}

\title{Partition Function Estimation: Quantum and Quantum-Inspired Algorithms}

\author{Andrew Jackson \orcid{0000-0002-5981-1604}}
\affiliation{Department of Physics, University of Warwick, Coventry CV4 7AL, United Kingdom}

\author{Theodoros Kapourniotis \orcid{0000-0002-6885-5916}}
\affiliation{Department of Physics, University of Warwick, Coventry CV4 7AL, United Kingdom}

\author{Animesh Datta \orcid{0000-0003-4021-4655}}
\affiliation{Department of Physics, University of Warwick, Coventry CV4 7AL, United Kingdom}

\date{\today}


\begin{abstract}
We present two algorithms, one quantum and one classical, for estimating partition functions of quantum spin Hamiltonians.
The former is a DQC1 (Deterministic quantum computation with one clean qubit) algorithm, 
and the first such for complex temperatures.
The latter, for real temperatures, achieves performance comparable to a state-of-the-art DQC1 algorithm [Chowdhury \emph{et al.} Phys. Rev. A 103, 032422 (2021)].
Both our algorithms take as input the Hamiltonian decomposed as a linear combination Pauli operators.
We show this decomposition to be DQC1-hard for a given Hamiltonian, providing new insight into the hardness of estimating partition functions. 

\end{abstract}

\maketitle
\emph{Introduction:}  
It is hoped that quantum computational devices will make more tractable classically intractable computations required for the study of condensed matter systems, including those in materials science, quantum chemistry~\cite{bauer2020quantum}, and farther afield~\cite{dodin2020applications}.
Meanwhile, classical computers remain the sole route to unravelling the properties of strongly correlated quantum systems
 such as the two-dimensional Hubbard model~\cite{2DH2015} or a chain of hydrogen atoms~\cite{Motta2017}.

Central amongst such problems is the computation of partition functions, which embody all thermodynamic information of a system at equilibrium.
Evaluating partition functions of spin systems exactly is \#P-hard~\cite{Jaeger1990} in the worst case~\cite{WelshChapter, Lidar_2004}.
Their exact evaluation is thus unlikely to be efficient with either classical or quantum devices.
The complexity barriers to exact evaluation are often circumvented by aiming for approximations, provided the error can be limited.
However, this can still be prohibitively hard~\cite{GoldbergGuo2017}.
Approximating the partition function for the classical Ising model with complex coefficients, even on the 2D square lattice, to exponentially large additive (absolute) error, 
is BQP-hard~\cite{2011QALGS, Matsuo_2014}.
This makes it unlikely to be tractable on classical computers, but may be possible on a BQP device \cite{10.1137/S0097539796300921, aaronson2009bqp}. 
Furthermore, for logarithmically-local quantum Hamiltonians, obtaining an approximation (of the partition function) to additive error exponential in both the size ($N$) of the system, and the product of the inverse temperature ($\beta$) with the sum of the lowest eigenvalue of each Hamiltonian term is DQC1-hard~\cite{chowdhury2019computing}.

DQC1 (Deterministic quantum computation with one clean qubit)~\cite{Knill_1998,Shepherd2006} is a complexity class believed to properly contain BPP but not be equivalent to BQP~\cite{10.5555/2017011.2017012,Aharonov_2008,DATTA2011}.
It is therefore not expected to be efficiently simulable classically to within multiplicative (relative) error unless the polynomial hierarchy collapses to the second level~\cite{Fujii_2018}.

Lately, the concepts of quantum computation have been deployed to develop quantum-inspired classical algorithms~\cite{Arrazola_2020}. 
Not only can these often unexpected results be useful in and of themselves~\cite{Tang_2019}, but also provide \textit{en passant} insight into the distinctions between classical and quantum computation~\cite{PhysRevLett.127.060503}. They also can be used more readily, rather than waiting for quantum technology to mature. 

\begin{table}
\begin{center}
\renewcommand{\arraystretch}{2.25}
\begin{tabular}{ | >{\centering}p{1.6cm} | >{\centering}p{1.9cm} | >{\centering}p{3.9cm} |}
\hline
Algorithm &			Computational model 			&	 Lower bound on expected runtime	\tabularnewline \hline
This Letter &	  Classical	 (BPP) & $\oo \left(\dfrac{ L^2 \beta^2 \xi^2 }{ \me^{2} \ln{(\me + 1)}}\left[ \dfrac{e^{L \beta\xi } 2^{N}}{Z} \right]^2 \right)$ \\ \tabularnewline \hline

Ref.~\cite{chowdhury2019computing} & Quantum (DQC1) &  $\oo \left( C_{\mathcal{H}} \dfrac{L^3 \beta^2 \xi^2 }{\me^{2} } \left[ \dfrac{ e^{L \beta \xi } 2^{N}}{ Z }  \right]^2 \right)$ \\ \tabularnewline \hline
\end{tabular}
\caption{Lower bound on the expected runtime of our quantum-inspired classical algorithm \emph{vis-a-vis} the DQC1 algorithm from Ref.~\cite{chowdhury2019computing} for approximating the partition function ($Z$) of a $N$-qubit Hamiltonian $\mathcal{H}$
consisting of $L$ terms with multiplicative error $\me$.
$\xi$ is the largest absolute value of the coefficients of the Pauli decomposition of $\mathcal{H}$
and $\beta$ is the inverse temperature.
Finally, $C_{\mathcal{H}} \sim \oo(\polylog(N)))$ is an overhead for implementing required unitaries in Ref.~\cite{chowdhury2019computing} using a quantum walk operator~\cite{Low_2019}.
The two runtimes being similar despite one algorithm being quantum and the other classical is not a contradiction, from a complexity perspective, as mapping from the input for Ref.~\cite{chowdhury2019computing} to the corresponding input for the algorithm presented herein is DQC1-hard.}
\label{AsympTable}
\end{center}
\end{table}

In this Letter, we present a quantum-inspired classical algorithm for approximating partition functions of quantum spin Hamiltonians at real temperatures 
to multiplicative error.
Its complexity is comparable (See Table \ref{AsympTable}) to a state-of-the-art DQC1 algorithm for the same problem~\cite{chowdhury2019computing}.
Along the way, we also present a DQC1 algorithm for the same task at complex temperatures to additive error.
The Hamiltonian input to our algorithms is expressed as a linear decomposition of Pauli operators with real coefficients.
Given a Hamiltonian as a set of circuits implementing each term and the associated coefficients (the form required for input to the algorithm in Ref. \cite{chowdhury2019computing}), we show the task of obtaining this decomposition to be DQC1-hard in the worst case.

Our Letter thus provides a new perspective on the hardness of estimating partition functions,
one requiring no mention of the sign problem~\cite{SignProblem}. It is rather based on the hardness of decomposing a Hamiltonian into a linear combination of tensor products of Pauli operators.
As the task of simulating quantum spin systems often begins with such a decomposition, 
their hardness may be dubbed the \textit{decomposition problem}.

\begin{figure}[h!]
\begin{center}
\begin{quantikz}[column sep=0.2cm]
\ket{0}	~ 		& \gate{H} 				& \ctrl{1} 					 & \gate{H}			  & \meter{} \\
\dfrac{\mathcal{I}}{2^N} ~	&	\qwbundle[alternate]{}	& \gate{U}  \qwbundle[alternate]{} & 	\qwbundle[alternate]{} & 	\qwbundle[alternate]{} \\ 
							\end{quantikz}
\end{center}
\caption{DQC1 circuit for estimating $\tr(U)/2^N$ for a $N$-qubit unitary $U.$ 
Its real and imaginary parts are obtained by measuring the $\mathcal{X}$ and $\mathcal{Y}$ Pauli operators respectively on the first qubit~\cite{Knill_1998}.}
\label{fig:TraceEstimation}
\end{figure}
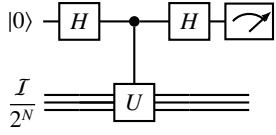

\emph{DQC1 Algorithm :}
The DQC1 model of quantum computation takes as inputs the completely mixed state, a single qubit in the state $\vert 0 \rangle$, and a description of a unitary $U$ as illustrated in Fig.~(\ref{fig:TraceEstimation}).
Estimating (via sampling) the expectation value of the first qubit with bounded error gives the output - an estimate of the normalised trace of the unitary operator $U$~\cite{Knill_1998}.

The partition function $Z$ for a  Hamiltonian $\mathcal{H}$ at a complex inverse temperature $\beta = b_R + ib_I, b_R, b_I \in \mathbb{R},$ is
given by
\be
\label{TraceDef}
   Z = \tr \left( \exp(- \beta \mathcal{H}) \right) = \tr \left(\exp( - b_R\mathcal{H}) \exp( - ib_I\mathcal{H}) \right).
\ee
Estimating the partition function using a DQC1 algorithm would require quantum circuits for the operators $\exp( - b_R\mathcal{H})$ and $\exp( - ib_I\mathcal{H}).$
These may be thought of as imaginary and real time evolutions, respectively.
We implement both approximately using the first-order Trotter decomposition. 
However, as $\exp( - b_R\mathcal{H})$ is not unitary, we cannot implement it directly and so have to implement it ``in the aggregate" by implementing various unitaries with different probabilities to implement it up to an easily calculated factor (see Lemma \ref{expZtheorem} for how this is done). This factor is corrected for by considering its effect on the DQC1 trace estimation algorithm (as in Fig. \ref{fig:TraceEstimation}).

Our DQC1 algorithm partition function estimation has two sources of errors -- from sampling and from the Trotter decomposition. 
Their origins lie in the trace estimation and the approximation by the first-order Trotter decomposition of the operators
 in Eq.~(\ref{TraceDef}) respectively, as outlined in the two preceding paragraphs.
Denoting these errors in the additive case as $\ase$ and $\ate,$ the total additive error in our algorithm is bounded by $\abse = \ase + \ate.$

Given an $\ate,$ the number of Trotter steps $\nu$ required to achieve it depends on the specifics of the problem.
Let  $\mathcal{H} =  \sum^L_{j=1} h_j \FP_j$ be a $N$-qubit Hamiltonian,
where $h_j \in \mathbb{R},$ each $\FP_j =  \mathcal{P}_1 \otimes \mathcal{P}_2 \otimes \cdots \otimes \mathcal{P}_n,$ 
acts on some subset of $n \leq N$ qubits labelled by $j$ and $ \mathcal{P}_j \in \{\mathcal{I}, \mathcal{X},  \mathcal{Y},  \mathcal{Z} \}$ is the set of single-qubit  operators.
Then the first-order Trotter decomposition of the imaginary time evolution is given by
\be
\label{eq:trotapp}
\mathcal{T}_R \approx \exp(-b_R \mathcal{H}), ~~\text{where} ~~ \mathcal{T}_R = \left[  \prod^L_{j=1}  \exp \left( c_j\FP_j \right) \right]^{\nu},  
\ee 
where $\nu$ is the number of Trotter steps and $c_j = - b_R h_j/\nu$.
$\mathcal{T}_I \approx \exp(-i b_I \mathcal{H})$ for the real time evolution is defined similarly with $b_R$ replaced by $i b_I$ (and we use the same number of Trotter steps, for convenience).
Trotter expansions have been well-studied for real time evolutions~\cite{Lloyd, Wecker2015, Heyleaau8342, benedetti2020hardwareefficient}.
In the following, we adapt the approach in Ref. \cite{Lloyd} for the imaginary time evolution.
We restrict ourselves to the first order given the dominance of gate errors over Trotter errors in near term devices~\cite{Knee_2015}.

These approximate operators lead to an approximation of the partition function
\be
\label{eq:troterr}
 Z \approx \zt \equiv \tr \left( \TR \TI \right)~~\text{ such that~~~~} | Z - \zt| \leq \ate.
 \ee
The number of Trotter steps $\nu$ required to achieve this Trotter error is given by Theorem \ref{TheoremTrotterComplex}.

\begin{theorem}
\label{TheoremTrotterComplex}
The number of Trotter steps $\nu$ required to implement the operator $\exp(-\beta \mathcal{H})$ using 
first-order Trotter decomposition with additive Trotter error, in the trace, $\ate$
is
    \be
    \label{eq:addtrot}
\nu = \oo \left(\frac{1}{\ate} 2^N |\beta|^2 \Omega^2  \exp(b_R \Omega) \right),
    \ee 
where   $\Omega  = \sum^L_{j=1} |h_j | . $
\end{theorem}

The proof is in Appendix \ref{DQC1Error}.

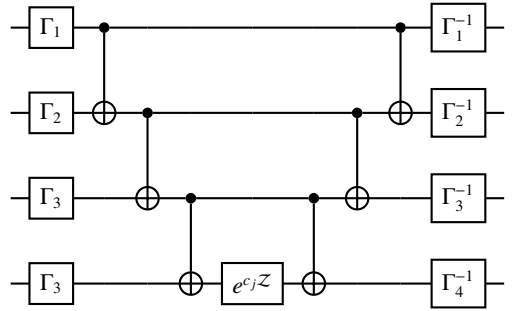
\begin{figure}[h!]
\begin{center}
\begin{quantikz}[column sep=0.25cm]
\qw 	& \gate{\Gamma_1}  & \ctrl{1}	& \qw 	& \qw 	&	\qw 					&	\qw 	& \qw 	& \ctrl{1} &  \gate{\Gamma_1^{-1}}  & \qw	\\
\qw 	& \gate{\Gamma_2}  & \targ{}	& \ctrl{1}    & \qw  	&	\qw 					&	\qw 	& \ctrl{1}	& \targ{} & \gate{\Gamma_2^{-1}}  & \qw	\\
\qw 	& \gate{\Gamma_3}  &	 \qw 		& \targ{}	& \ctrl{1} 	& \qw 				& \ctrl{1}   & \targ{} 	& \qw	&  \gate{\Gamma_3^{-1}}  &\qw		\\
\qw 	& \gate{\Gamma_3}  &	 \qw 		&  \qw 	& \targ{}	& \gate{e^{c_j\mathcal{Z}}} & \targ{}  & \qw		& \qw	& \gate{\Gamma_4^{-1}}  & \qw		
\end{quantikz}
\end{center}
\caption{Gadget implementing the imaginary time (non-unitary) evolution 
$\exp(-b_R \mathcal{H})$ on $n=4$ qubits. $\Gamma_j \in \{H,R_x(\pi/2)\}.$ See Appendix \ref{imagtimeEvol} for details.}
    \label{fig:nonUnitGadget}
\end{figure}

Appendix~\ref{imagtimeEvol} shows how each of these Trotter steps can be implemented via the gadget in Fig.~(\ref{fig:nonUnitGadget}) (for the imaginary time) or as in Ref. \cite{Lloyd} (for real time).
It uses Lemma \ref{expZtheorem} which holds because the Pauli operators square to the identity, and is proved in Appendix \ref{expZtheoremproof}.

\begin{lemma}
\label{expZtheorem}
Let $c \in \mathbb{R}$ and $\FP$ ($\FI$) be the tensor product of single-qubit Pauli, including identities, (identity) operators on any number of qubits. Then
\be
\label{re-expression}
   \exp(c \FP)    = (\cosh c) \mathcal{I} + \sinh \vert c \vert \left( \Sigma(c) \FP \right),
  \ee
where $\Sigma(c)$ denotes the sign of $c$.
\end{lemma}

As $\cosh c/\!\exp|c|$ and $\sinh|c|/\!\exp|c|$ are positive and add to unity, Lemma \ref{expZtheorem} 
enables implementing each $\exp(c_j \FP_j)$ in Eq.~(\ref{eq:trotapp}) probabilistically, up to the constant $\exp|c_j|$.
Indeed, the quantum circuits corresponding to both the terms are reducible to single-qubit ones, as illustrated in Fig.~(\ref{fig:GadgetZI}). 
The sign $\Sigma(c_j)$ associated with $c_j$ is omitted there, and is applied as a phase when implementing the Pauli (as the entire implementation of $\exp(c \FP)$ is controlled this is not an overall phase).
This probabilistic implementation of the non-unitary evolution contributes to the sampling error in our algorithm.

In particular, given an $\ase,$ the number of samples $\nu_{\mathrm{S}}$ (or repetitions of the circuit in Fig.~(\ref{fig:TraceEstimation})) required to achieve it is
\be
\label{eq:addsam}
\nu_{\mathrm{S}} = \oo \left(\frac{1}{ \ase^{2}}2^{2N} \exp(2|b_R|\Omega) \right).
\ee

The runtime of our DQC1 algorithm is given by the product of the bounds in Eqs.~(\ref{eq:addtrot}) and (\ref{eq:addsam}).

\begin{figure}[h!]
\begin{center}

\begin{quantikz}[column sep=0.15cm]
\qw 	& \gate{\Gamma_1}  & \ctrl{1}	& \qw 	& \qw 	&	\qw 					&	\qw 	& \qw 	& \ctrl{1} &  \gate{\Gamma_1^{-1}}  & \qw	\\
\qw 	& \gate{\Gamma_2}  & \targ{}	& \ctrl{1}    & \qw  	&	\qw 					&	\qw 	& \ctrl{1}	& \targ{} & \gate{\Gamma_2^{-1}}  & \qw	\\
\qw 	& \gate{\Gamma_3}  &	 \qw 		& \targ{}	& \ctrl{1} 	& \qw 				& \ctrl{1}   & \targ{} 	& \qw	&  \gate{\Gamma_3^{-1}}  &\qw		\\
\qw 	& \gate{\Gamma_3}  &	 \qw 		&  \qw 	& \targ{}	& \gate{\mathcal{I}} & \targ{}  & \qw		& \qw	& \gate{\Gamma_4^{-1}}  & \qw		
\end{quantikz}
=
\begin{quantikz} [row sep=0.85cm]
	& \ghost{} & \qw & \qw  & \qw  & \qw \\
	& \ghost{}  & \qw & \qw  & \qw  & \qw\\
	& \ghost{}  & \qw & \qw  & \qw  & \qw\\
	& \ghost{}   & \qw & \qw  & \qw  & \qw
\end{quantikz}
\begin{quantikz}[column sep=0.15cm]
\qw 	& \gate{\Gamma_1}  & \ctrl{1}	& \qw 	& \qw 	&	\qw 					&	\qw 	& \qw 	& \ctrl{1} &  \gate{\Gamma_1^{-1}}  & \qw	\\
\qw 	& \gate{\Gamma_2}  & \targ{}	& \ctrl{1}    & \qw  	&	\qw 					&	\qw 	& \ctrl{1}	& \targ{} & \gate{\Gamma_2^{-1}}  & \qw	\\
\qw 	& \gate{\Gamma_3}  &	 \qw 		& \targ{}	& \ctrl{1} 	& \qw 				& \ctrl{1}   & \targ{} 	& \qw	&  \gate{\Gamma_3^{-1}}  &\qw		\\
\qw 	& \gate{\Gamma_3}  &	 \qw 		&  \qw 	& \targ{}	& \gate{\mathcal{Z}} & \targ{}  & \qw		& \qw	& \gate{\Gamma_4^{-1}}  & \qw		
\end{quantikz}
~~=
\begin{quantikz}
	& \gate{\Gamma_1 \mathcal{Z} \Gamma_1^{-1}} & \qw\\
	& \gate{\Gamma_2 \mathcal{Z} \Gamma_2^{-1}} & \qw\\
	& \gate{\Gamma_3 \mathcal{Z} \Gamma_3^{-1}} & \qw\\
	& \gate{\Gamma_4 \mathcal{Z} \Gamma_4^{-1}}  & \qw
\end{quantikz} 
\end{center}
\caption{Quantum circuits corresponding to the two terms in Eq.~(\ref{re-expression}).}
    \label{fig:GadgetZI}
\end{figure}

\emph{Classical Algorithm:}
Our quantum-inspired classical algorithm differs in two respects from the preceding DQC1 algorithm. 
It takes a \emph{real} inverse temperature $\beta \in \mathbb{R}$ as input, and outputs an estimate of the partition function 
\begin{align}
\label{TraceDefreal}
   Z = \tr \left( \exp(- \beta \mathcal{H}) \right).
\end{align}
to \emph{multiplicative} error.
This classical algorithm is formally stated as Algorithm~\ref{algorithmFormal} in Appendix \ref{sampBoundProof}.
The classical algorithm is essentially the same as the aforementioned DQC1 algorithm but the insistence on the inverse temperature being real now means all the gates in Figs.~(\ref{fig:TraceEstimation}) and (\ref{fig:GadgetZI}) are Clifford (for our purposes, it's most important that the right hand side of Fig. (\ref{fig:GadgetZI}) consists entirely of Pauli gates as $\forall k \in \mathbb{N}$, $\Gamma_k \mathcal{Z} \Gamma_k^{-1}$ reduces to a single Pauli gate, via Lemma \ref{Twirly}).
Consequently, the right hand side of Fig. (\ref{fig:GadgetZI}) can be implemented classically~\cite{gottesman1998heisenberg, AaronsonGottesman2004} and so can a controlled version of  Fig. (\ref{fig:GadgetZI}) (as required for our algorithm).
Again, there are two sources of errors -- from the Trotter decomposition and from sampling. 

We consider the Trotter error first. 
If the multiplicative Trotter error in approximating $Z$ by $\zt = \tr \left( \mathcal{T}_R  \right) $ is given by $\mte,$ then
\be
\label{eq:multerr}
| Z - \zt  | \leq Z \mte,
\ee
and the number of Trotter steps $\nu$ is determined by Theorem \ref{TraceErrorTheorem}.
Its proof is in Appendix~\ref{errorBoundProof}, and uses methods from Ref.~\cite{childs2019theory}.

To fully exploit mutual non-commutativity amongst the terms of $\mathcal{H},$ we require the notion of the non-commuting set~\cite{childs2019theory}.

\begin{definition}[$k$'th non-commuting set]
\label{Non-commuteSet}
    
The $k$'th non-commuting set $\zeta_k$ is the set of all terms in the Hamiltonian $\mathcal{H} =  \sum^L_{j=1} h_j \FP_j$ that do not commute with the $k$'th term of the Hamiltonian $\FP_k$.
\end{definition}

\begin{theorem}
\label{TraceErrorTheorem}
The number of Trotter steps $\nu$ required to approximate $Z = \tr \left( \exp(- \beta \mathcal{H}) \right)$ up to multiplicative error $\mte$
using the first-order Trotter decomposition 
is
\be
\label{eq:TrotNum}
\nu \geq \frac{ \beta^2 \Omega \mathfrak{h}}{\ln(1+\mte)},
\ee
where $\Omega =  \sum_{j = 1}^{L} |h_j| $ and $\mathfrak{h} = \sum^L_{k=2} |h_k| N_k$ with
$N_k = |\{z \leq k \text{~such that~} z \in \mathcal{\zeta}_k \} |.$ 
\end{theorem}

For the sampling error, we follow  Ref.~\cite[Algorithm 3]{chowdhury2019computing} (For completeness, we present it as Algorithm~\ref{algorithmFormal} in Appendix \ref{sampBoundProof}.).
It shows that estimates with successively smaller additive error lead to an estimate to within the required multiplicative error, with probability as high as required. 
If the multiplicative sampling error is $\mse$ and $Z_{\mathrm{max}}$ an upper bound on the value of the partition function,
then this algorithm provides an estimate $Z_{\mathrm{S}}$ of $\zt$ in Eq.~\ref{eq:multerr} such that
\be
\label{eq:mse}
\mathrm{Pr}\left(| \zt - Z_{\mathrm{S}}  | \leq \zt  \mse \right) \geq 1 - \delta,
\ee
where $\delta >0$ is the upper bound on the probability of obtaining an estimate beyond the precision $\mse.$
 The runtime of this algorithm is a random variable with an expected value of~\cite{chowdhury2019computing}
 \be
 \label{eq:exptime}
T_{\mathrm{S}} = \oo\left(\frac{2^{2N} \exp(2\beta \Omega) }{ \mse^2 Z^2} \log_2 \left(\frac{1}{\delta} \right)\log_2 \left(\frac{Z_{\mathrm{max}}}{\zt} \right) \right).
 \ee

The multiplicative Trotter and sampling errors in our estimation algorithm for the partition function combine to give a 
total multiplicative error $\me$ such that 
\be
\me  = \mse + \mte + \mse\mte.
\ee
This follows from Eqs.~(\ref{eq:multerr}) and (\ref{eq:mse}), the positivity of the partition functions, their estimates, and the errors.
Using this in conjunction with Eqns.~(\ref{eq:TrotNum}), (\ref{eq:exptime}) and neglecting the $\log_2(\cdot)$ contributions 
gives the total expected runtime of our classical algorithm as in Table~\ref{AsympTable}. 
It is comparable to that of the DQC1 algorithm in Ref.~\cite{chowdhury2019computing}, 
despite the belief that DQC1 is more powerful than classical computation~\cite{Knill_1998,Shepherd2006,10.5555/2017011.2017012,Aharonov_2008,DATTA2011}.
This is because the two algorithms take as input the Hamiltonian in two different forms which encode different degrees of hardness, as we show next.

\emph{Hardness of Partition Function Estimation}: 
Our classical algorithm requires the Hamiltonian input as a linear combination of tensor products of Pauli operators.
This is a special case of the Hamiltonian input as a linear combination of unitary operators, 
each of which can be implemented by an efficient quantum circuit, which is the input format Ref. \cite{chowdhury2019computing} requires.
In Appendix~\ref{HardnessProof}, we show that given a Hamiltonian, as a linear combination of unitary operators with the associated circuit,  obtaining the same Hamiltonian as a linear combination of tensor products of Pauli operators (what our algorithm takes as input) is DQC1-hard.
This shows that part of the hardness of estimating partition functions of quantum spin Hamiltonians may be ascribed to a \textit{decomposition problem}.

\emph{Numerical Results}: 
As the purpose of an algorithm is to solve a problem, we now present results of numerical investigations into our classical partition function estimation algorithm 
(Algorithm~\ref{algorithmFormal} in Appendix \ref{sampBoundProof}).
To recapitulate, our classical algorithm is designed for real temperatures and estimates the partition function up to a multiplicative  error.

\begin{figure}[h!]
    \centering
    \includegraphics[width=0.5\textwidth]{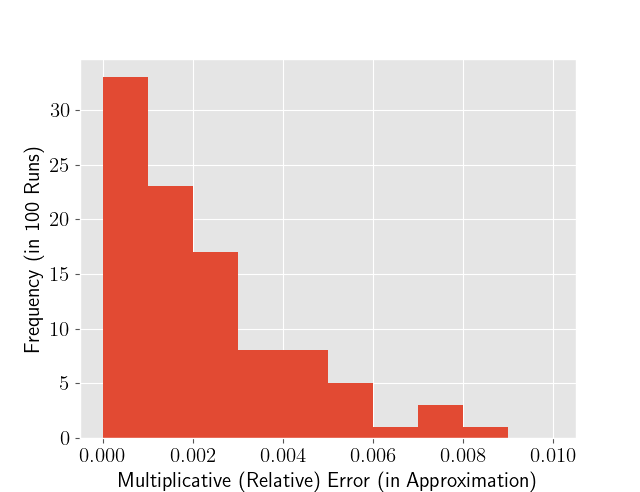}
    \caption{Histogram of the multiplicative (relative) error in our algorithm's estimation of 100 randomly generated Hamiltonians on between 1 and 3 spins.}
    \label{CorrectnessTestInPaper}
\end{figure}

We first experimentally verify the correctness of our Algorithm~\ref{algorithmFormal}, although it follows formally from Ref.~\cite{chowdhury2019computing}.
To that end, we generate 100 Hamiltonians with (uniformly in [-1,1]) random coefficients $h_j$ of up to $L=4$ random Pauli terms each acting on up to $N=3$ spins.
We then estimate their partition functions at random real inverse temperatures $\beta $ using our Algorithm~\ref{algorithmFormal} 
for $\mse = \mte = 0.048, \delta = 0.15, $ and $ Z_{\mathrm{max}} $ was set to twice the true value of the partition function.
The latter was obtained, up to numerical error, using full diagonalisation.
Comparing the estimate and the true values shows that our Algorithm~\ref{algorithmFormal} indeed produces estimates well within $\me \approx 0.098$ of the exact value,
as shown in Fig.~\ref{CorrectnessTestInPaper}.

To illustrate the performance of our algorithm in a problem of interest, we resort to the  two-dimensional (2D) one-band Fermi-Hubbard model, 
solutions of which have been widely studied using different numerical algorithms~\cite{2DH2015}.
Its simplest rendition is given by
\be
\label{eq:2dhib}
    \mathcal{H} = - t \!\!\!  \!\!\!  \sum_{\scriptsize{
\begin{array}{c}
\langle i,j \rangle \\
\sigma \in \{\uparrow, \downarrow \}
\end{array}
}}
   \!\!\!  \!\!\! 
     \left( \hat{C}^{\dag}_{i, \sigma} \hat{C}_{j, \sigma} + \hat{C}^{\dag}_{j, \sigma} \hat{C}_{i, \sigma} \right) 
     + U \!\!  \sum_{k}  \hat{n}_{k, \uparrow} \hat{n}_{k, \downarrow},
\ee
where $\langle i,j \rangle$ indicates adjacent vertices $i,j$ of a 2D graph,
$\hat{C}^{\dag}_{i, \sigma},\hat{C}_{i, \sigma}$ and $\hat{n}_{i, \sigma} = \hat{C}^{\dag}_{i, \sigma} \hat{C}_{i, \sigma}$ 
denote Fermionic creation, annihilation and number operators respectively, for spin $\sigma$ at vertex $i$ of the graph.
$t$ denotes the nearest-neighbour hopping strength and $U$ the onsite interaction strength.
It is typical to set the energy scales in the Hamiltonian in terms of $t.$ 
Thus, the dimensionless onsite interaction strength is given by $\tilde{U} = U/t$ and the dimensionless inverse temperature by $\tilde{\beta} = \beta t.$

\begin{figure}[h!]
    \centering
    \includegraphics[width=0.5\textwidth]{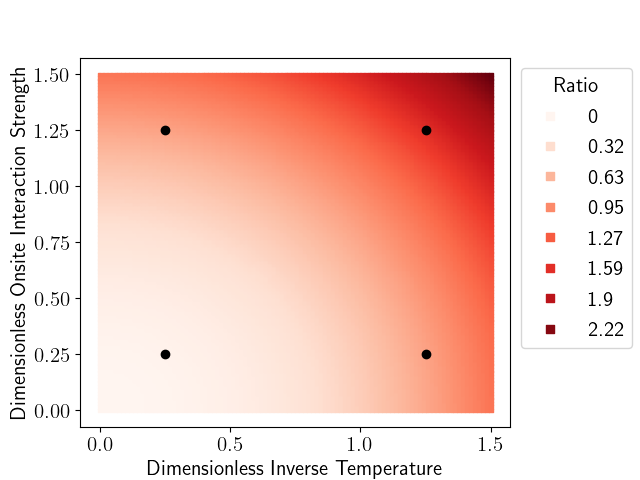}
    \caption{ Numerical Test of Time Requirements.
    Darker red indicates a greater value of the ratio of the time our algorithms takes compared to full diagonalisation using ALPS.
    The black dots indicate the value is given in Table \ref{ValuesTable}.
    These numbers were obtained using a Dell PowerEdge C6420 with 2 x Intel Xeon Platinum 8268 (Cascade Lake) 2.9 GHz 24-core processors with 192 GB of (DDR4-2933) RAM.}
    \label{PerformanceHeatmap}
\end{figure}

Being inspired by a quantum algorithm (Ref.~\cite{chowdhury2019computing}), the above Hamiltonian must undergo a Fermion to qubit mapping before being fed into our Algorithm~\ref{algorithmFormal}.
We use a recent low-weight encoding~\cite{derby2020low} for this.
For a $3 \times 3$ square lattice, this leads to a Hamiltonian on $N=26$ qubits with $L=60$ terms.
The projected ratio between the time to estimate the partition function using our Algorithm~\ref{algorithmFormal} 
(with $\mse = \mte = 0.15, \delta = 0.20$) and the time to estimate using full diagonalisation in ALPS \footnote{Algorithms and Libraries for Physics Simulations, Version 2.3. Available at \url{http://alps.comp-phys.org/mediawiki/index.php/Main_Page}.}, is shown in Fig.~\ref{PerformanceHeatmap}.
This enables a comparison of the performance of our Algorithm~\ref{algorithmFormal} against a recognised benchmark, as illustrated in Table~\ref{ValuesTable}.

Unsurprisingly, in Fig.~\ref{PerformanceHeatmap} and Table~\ref{ValuesTable}, larger $\tilde{\beta} $ and $\tilde{U} $ results in our algorithm performing comparatively worse.
Furthermore, an increase in $\tilde{\beta} $ has a larger effect than an equivalent one in $\tilde{U} $.
The latter follows from Eq.~(\ref{eq:TrotNum}) where an increase in $\tilde{U} $ only increases some of the terms in $\Omega$ and $\mathfrak{h},$
whereas an equivalent increase in $\tilde{\beta} $ is tantamount to increasing all the terms in $\Omega$ and $\mathfrak{h}.$

As to the space requirements, our algorithm requires memory scaling linearly in $N.$  
This is favourable compared to full diagonalisation which requires processing matrices exponentially large in $N$ and is thus very memory-intensive.

\begin{table}[t!]
\begin{center}
\renewcommand{\arraystretch}{2.0}
\begin{tabular}{ | >{\centering}p{1.0cm} | >{\centering}p{3.5cm} | >{\centering}p{3.5cm} |  }
\hline
 \diagbox[width=1.0cm, height=1cm]{$\tilde{U}$}{\raisebox{0.0ex}{$\tilde{\beta}$}} &	$0.25$			&	 $1.25$ 	\tabularnewline \hline
			$1.25$ 	         	& $0.0522$	  				&   $1.3043$ \tabularnewline \hline
			$0.25$	                 & $0.0215$	 & $0.5351$	 \tabularnewline \hline
\end{tabular}
\caption{The ratio of the time our algorithms takes compared to full diagonalisation using ALPS for the four marked points in Fig.~\ref{PerformanceHeatmap}.}
\label{ValuesTable}
\end{center}
\end{table}

\emph{Conclusions}:
Our Letter provides a different perspective on the root of the hardness of estimating partition functions. 
In this view, the hardness of estimating partition functions may be due to the \textit{decomposition problem}.
This vantage may provide useful insights into the complexity of simulating hard quantum problems, as has been the case for the sign problem~\cite{Hen2021,Berger_2021}.
For instance, the exponential contributions in our runtime (Table~\ref{AsympTable}) arise from sampling (Eq.~\ref{eq:exptime}).
These, and the runtime as a whole are independent of the sign of the Hamiltonian in certain bases. 
Thus our classical algorithm may be exponential even for sign-problem-free systems. 
However, it could be significantly more efficient when the partition function is large.
Low temperatures, of typical interest, do not correspond to such a scenario.
Endeavours to explore other Hamiltonians and parameters which do could be worthwhile, 
given the general hardness and importance of estimating partition functions.

\emph{Acknowledgements}:
We are grateful to Andreas Honecker and David Quigley for sharing their expertise.
This work was supported, in part, by the UK Networked Quantum Information Technologies (NQIT) Hub (EP/M013243/1), 
the UKRI ExCALIBUR project QEVEC (EP/W00772X/2), 
and a Leverhulme Trust Early Career Fellowship.
We acknowledge the use of the Scientific Computing Research Technology Platform at the University of Warwick.

\bibliography{References}

\newpage
\appendix 
\begin{widetext}

\section{Proof of Theorem \ref{TheoremTrotterComplex}}
\label{DQC1Error}
Following the main text, let $\TR, \TI$ be the first-order Trotter approximations to $e^{-b_R \mathcal{H}}$, $e^{-ib_I \mathcal{H}}$ respectively with $\nu$ Trotter steps.
Denote
\be
    \Delta  =     \bigg \vert \tr \left( e^{-\beta \mathcal{H}} \right) - \tr \left( \TR \TI \right) \bigg \vert.
\ee
Then,
\begin{align}
    \Delta
    &= 
    \bigg \vert \tr \left( e^{-\beta \mathcal{H}} \right) - \tr \left( \TR \TI \right) + \tr \left( e^{-b_R \mathcal{H}} \TI  \right) - \tr \left( e^{-b_R \mathcal{H}} \TI  \right)\bigg \vert
    =
    \left \vert \tr \left( e^{-b_R \mathcal{H}} \left [ e^{-ib_I \mathcal{H}}  - \TI \right ]\right)  +  \tr \left( \left[ -\TR + e^{-b_R \mathcal{H}} \right ] \TI \right) \right \vert \\
    &\leq
    \left \vert \tr \left( e^{-b_R \mathcal{H}} \left [ e^{-ib_I \mathcal{H}}  - \TI \right ]\right) \right \vert + \left \vert \tr \left( \left[ \TR - e^{-b_R \mathcal{H}} \right ] \TI \right) \right \vert
\end{align}
Denoting $\vert \vert \cdot \vert \vert$ as the spectral norm -  the largest singular value,
\be
 \Delta
	\leq           2^N \SN{ e^{-b_R \mathcal{H}} \left [ e^{-ib_I \mathcal{H}}  - \TI \right ]} +2^N \SN{ \left[ \TR - e^{-b_R \mathcal{H}} \right ] \TI }
	\leq           2^N \SN{ e^{-b_R \mathcal{H}}} \SN{ e^{-ib_I \mathcal{H}}  - \TI } +2^N \SN{  \TR - e^{-b_R \mathcal{H}}} ,
\label{eq:errbnd}
\ee   
where we have used the sub-multiplicative property of the spectral norm and $\SN{ \TI }=1$ as $\TI$ is a unitary. 
Lemma~\ref{RequiredForBounding} shows that
\ben
    \SN{ \exp(-ib_I \mathcal{H})  - \TI } &\leq& \nu \SN{   \exp\left(- i \frac{b_I}{\nu} \mathcal{H} \right) - \oTI}, \\
    \SN{  \TR - \exp (-b_R \mathcal{H})}  &\leq& \nu \exp \left( \left(1 -\frac{1}{\nu} \right) |b_R| \Omega   \right)\SN{  \oTR - \exp\left(-\frac{b_R}{\nu} \mathcal{H} \right)}  , 
\een
where $\oTR = [\TR]^{1/\nu} = \prod^L_{j=1}  \exp \left( c_j\FP_j \right) $ denotes one Trotter step for the imaginary time evolution, 
$\oTI$ similarly denotes one Trotter step for the real time evolution, and
$\Omega =  \sum^L_{j=1} \SN{h_j \FP} =  \sum^L_{j=1} |h_j | . $

Using these with Lemma 1 in Ref.~\cite{childs2019theory} for the first-order Trotter expansion,
\be
    \SN{  \TR - \exp (-b_R \mathcal{H})} = \oo \left( \frac{b_R^2}{\nu} \Omega^2 \exp(b_R \Omega)\right)
    ~~~\text{and}~~~
    \SN{ \exp(-ib_I \mathcal{H})  - \TI } = \oo \left( \frac{b_I^2}{\nu} \Omega^2 \right).
\ee
Substituting these in Eq.~(\ref{eq:errbnd}), and using Eq.~(\ref{eq:ebr}) 
\be
    \Delta \leq \oo \left(\frac{1}{\nu} 2^N \Omega^2 |\beta|^2 \exp(b_R \Omega) \right)	\leq \ate,
\ee
where the last inequality uses Eq.~(\ref{eq:troterr}).
Thus,
\be
    \nu \geq \oo \left(\frac{1}{\ate} 2^N \Omega^2 |\beta|^2 \exp(b_R \Omega) \right)	.
\ee

\begin{lemma}
\label{RequiredForBounding}
For a Hamiltonian $\mathcal{H} =  \sum^L_{j=1} h_j \FP$ and $b_R, b_I \in \mathbb{R},$ 
\ben
    \SN{  \TR - \exp (-b_R \mathcal{H})}  &\leq& \nu \exp \left( \left(1 -\frac{1}{\nu} \right) |b_R| \Omega   \right)\SN{  \oTR - \exp\left(-\frac{b_R}{\nu} \mathcal{H} \right)}  , \\
    \SN{ \exp(-ib_I \mathcal{H})  - \TI } &\leq& \nu \SN{   \exp\left(- i \frac{b_I}{\nu} \mathcal{H} \right) - \oTI},
\een
where $\Omega =  \sum^L_{j=1} \SN{h_j \FP} =  \sum^L_{j=1} |h_j | . $
\end{lemma}

\begin{proof} 
Using the triangle inequality and sub-multiplicativity of the spectral norm, 
\begin{align}
    \bigg \vert \bigg \vert \TR - \exp(-b_R \mathcal{H}) \bigg \vert \bigg \vert
    &=
    \bigg \vert \bigg \vert  \left[\oTR\right]^{\nu} - \exp(-b_R \mathcal{H}) \bigg \vert \bigg \vert\\
    &=
    \bigg \vert \bigg \vert
    \left[\oTR\right]^{\nu}
    -
    \exp\left(-\frac{b_R}{\nu} \mathcal{H}\right)  \left[\oTR\right]^{\nu-1}
    +
    \exp\left(-\frac{b_R}{\nu} \mathcal{H}\right)  \left[\oTR\right]^{\nu-1} 
     -
    \exp(-b_R \mathcal{H})
    \bigg \vert \bigg \vert\\
    &\leq
    \bigg \vert \bigg \vert
    \left[\oTR\right]^{\nu}
    -
     \exp\left(-\frac{b_R}{\nu} \mathcal{H}\right)  \left[\oTR\right]^{\nu-1}  \bigg \vert \bigg \vert
    +  
      \bigg \vert \bigg \vert
   \exp\left(-\frac{b_R}{\nu} \mathcal{H}\right)  \left[\oTR\right]^{\nu-1} 
     -
    \exp(-b_R \mathcal{H})
    \bigg \vert \bigg \vert\\
    &\leq
    \label{LastInFirstBlock}
      \SN{ \left[\oTR\right]^{\nu-1}} 
      \SN{ \oTR   -   \exp\left(-\frac{b_R}{\nu} \mathcal{H}\right) } 
      + \SN{ \exp\left(-\frac{b_R}{\nu} \mathcal{H}\right)}
      \SN{ \left[\oTR\right]^{\nu-1} - \exp\left(-b_R \left(1 - \frac{1}{\nu} \right) \mathcal{H} \right)}  .
\end{align}
Recursively applying the above procedure for the second term,
\begin{align}
    \SN{\TR - \exp(-b_R \mathcal{H}) }
   \leq
    \left(   \sum_{\substack{j, k \geq 0\\ j + k = \nu-1}} 
     \SN{\exp\left(-\frac{b_R}{\nu} \mathcal{H}\right)}^j  \SN{\oTR}^k \right)
    \bigg \vert \bigg \vert
    \oTR    -    \exp \left( -\frac{b_R}{\nu} \mathcal{H} \right)
    \bigg \vert \bigg \vert.
    \label{eq:jknu}
\end{align}
Since $ \SN{ \FP }=1$ and $c_j = - b_R h_j/\nu,$
\be
    \SN{\oTR} 
    = \SN{\prod^L_{j=1}  \exp \left( c_j\FP \right) }
    \leq \prod^L_{j=1} \SN{ \exp \left( c_j\FP \right)}
    \leq \prod^L_{j=1}  \exp \left( |c_j |\right)
    = \exp \left( \sum^L_{j=1} |c_j |\right)
    =  \exp \left(\frac{|b_R|}{\nu} \Omega \right),
\ee
and
\be
    \label{eq:ebr}
    \SN{\exp\left(-\frac{b_R}{\nu} \mathcal{H}\right)} \leq \exp \left(\frac{|b_R|}{\nu} \Omega \right).
\ee
Using the above bounds on each of the terms in the sum of Eq.~(\ref{eq:jknu}) gives the final result.
The result for the real time evolution follows similarly, albeit more simply as $\SN{\oTI}=1 =  \SN{\exp\left(-i \frac{b_I}{\nu} \mathcal{H}\right)}.$
\end{proof}

\section{Implementing Imaginary Time Evolution}
\label{imagtimeEvol}
In this Appendix, we show that gadgets in Fig.~(\ref{fig:nonUnitGadget}), analogous to those in Ref.~\cite{Lloyd}, can be used to implement
imaginary time evolutions such as $\exp(b_R \mathcal{H}).$

\subsection{Overview and preparatory lemmas}
We denote by $ \mathcal{P} = \left \{ \mathcal{X},  \mathcal{Y},  \mathcal{Z} \right\}$ the set of single-qubit Pauli operators,
by $ \mathcal{P}_j \in \mathcal{P}$ a Pauli operator on qubit $j,$ and
by $\Gamma_j \in \{H,R_x(\pi/2)\}$ the single-qubit Hadamard gate and that which implements a rotation about the $x$ axis by $\pi.$ 
Then
\begin{lemma}
\label{Twirly}
$   
    \mathcal{X} = H \mathcal{Z} H$ and $ \mathcal{Y} = R_x(\pi/2) \mathcal{Z}R_x(-\pi/2)
$
\end{lemma}
\begin{proof}
    Follows from matrix multiplication.
\end{proof}

\begin{lemma} 
\label{Lem:2to1}
For any real or complex number $h_j,$ given a gadget for the time evolution of the Hamiltonian
$
    \mathcal{H}_1 = h_j \cdot\mathcal{Z}_1 \otimes\mathcal{Z}_2 \otimes\mathcal{Z}_3 \otimes \cdots \otimes\mathcal{Z}_n,
$
on $n$ qubits, the time evolution of any Hamiltonian 
$
    \mathcal{H}_2 = h_j \cdot \mathcal{P}_1 \otimes \mathcal{P}_2 \otimes \cdots \otimes \mathcal{P}_n  
$
of Pauli operators may be implemented through the addition of the single-qubit gates $\Gamma.$
\end{lemma}

\begin{proof}
As each Pauli operator acts on a different qubit, they mutually commute, giving
\be
    e^{\mathcal{H}_2} 
    =  e^{ h_j \cdot \mathcal{P}_1 \otimes \mathcal{P}_2 \otimes \cdots \otimes \mathcal{P}_n }
    = \sum^{\infty}_{k=0} \bigg( \dfrac{\big( h_j \cdot \mathcal{P}_1 \mathcal{P}_2 \cdots \mathcal{P}_n \big)^k}{k!} \bigg) 
    = \sum^{\infty}_{k=0} \bigg( \dfrac{\big( h_j\big)^k \big(\mathcal{P}_1 \big)^k \big( \mathcal{P}_2 \big)^k \cdots \big(\mathcal{P}_n\big)^k}{k!} \bigg),
\ee
where we have suppressed the $\otimes$ for brevity.
Using Lemma~(\ref{Twirly}),
$    
    \mathcal{P}_i = \Gamma_i\mathcal{Z}_i \Gamma^{-1}_i,
$
gives
\ben
    e^{\mathcal{H}_2} 
    &=& \sum^{\infty}_{k=0} \bigg( \dfrac{\big(h_j \big)^k \big( \Gamma_1\mathcal{Z}_1 \Gamma^{-1}_1 \big)^k \cdots \big(\Gamma_n\mathcal{Z}_n \Gamma^{-1}_n\big)^k}{k!} \bigg)
    = \sum^{\infty}_{k=0} \bigg( \dfrac{\big( h_j \big)^k  \Gamma_1 \big(Z_1\big)^k \Gamma^{-1}_1 \cdots \Gamma_n \big(\mathcal{Z}_n \big)^k\Gamma^{-1}_n}{k!} \bigg)\\
    &=&  \sum^{\infty}_{k=0} \bigg( \Gamma_1 \cdots \Gamma_n \dfrac{\big( h_j \big)^k  \big(Z_1\big)^k \cdots \big(\mathcal{Z}_n \big)^k}{k!} \Gamma^{-1}_1 \cdots \Gamma^{-1}_n\bigg)
    = \Gamma_1 \cdots \Gamma_n \sum^{\infty}_{k=0} \bigg( \dfrac{\big( h_j \cdot\mathcal{Z}_1 \cdots\mathcal{Z}_n  \big)^k}{k!}\bigg)\Gamma^{-1}_1\cdots \Gamma^{-1}_n
    = \left(\Gamma_1 \cdots \Gamma_n \right) e^{\mathcal{H}_1} \left( \Gamma^{-1}_1 \cdots \Gamma^{-1}_n \right) \nonumber,
\een
where each $\Gamma_j$ can be identified and implemented efficiently a single gate.
\end{proof}

To obtain a quantum circuit for implementing $e^{\mathcal{H}_2},$ 
we begin with a quantum circuit for implementing $e^{\mathcal{H}_1}.$

\subsection{Implementing $e^{\mathcal{H}_1}$}
We begin with a quantum circuit for implementing $e^{\mathcal{H}_1}$ for just two qubits. 

\begin{lemma}
The quantum circuit in Fig.~($\ref{fig:ExampleHamCirc}$) implements $e^{\mathcal{H}}$ where $\mathcal{H} = h_j \cdot\mathcal{Z}_1 \otimes\mathcal{Z}_2$.

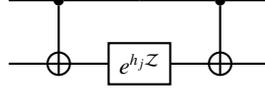
\begin{figure}[h!]
\begin{center}
\begin{quantikz}
& \ctrl{1}	& \qw & \ctrl{1} & \qw\\
& \targ{}	& \gate{e^{h_j\mathcal{Z}}} & \targ{} & \qw
\end{quantikz}
\end{center}
\caption{Quantum circuit to implement $e^{\mathcal{H}}$ where $\mathcal{H} = h_j \cdot\mathcal{Z}_1 \otimes\mathcal{Z}_2.$}
 \label{fig:ExampleHamCirc}
\end{figure}
\end{lemma}

\begin{proof} 
Considering the matrix representation of the circuit in the computational basis gives
\be
    \begin{bmatrix} 1 & 0 & 0 & 0 \\ 0 & 1 & 0 & 0 \\ 0 & 0 & 0 & 1\\ 0 & 0 & 1 & 0\end{bmatrix} 
    \left(\mathcal{I} \otimes \begin{bmatrix}e^{h_j} & 0 \\ 0 & e^{-h_j} \end{bmatrix} \right)
    \begin{bmatrix} 1 & 0 & 0 & 0 \\ 0 & 1 & 0 & 0 \\ 0 & 0 & 0 & 1\\ 0 & 0 & 1 & 0\end{bmatrix}
    = \begin{bmatrix} e^{h_j} & 0 & 0 & 0\\ 0 & e^{-h_j} & 0 & 0\\ 0 & 0 & e^{-h_j} & 0\\ 0 & 0 & 0 & e^{h_j} \end{bmatrix}
\ee
where $\mathcal{I}$ denotes the identity gate.
This is identical to 
\be
    e^{ \mathcal{H}}
    = \begin{bmatrix} e^{h_j} & 0 & 0 & 0\\ 0 & e^{-h_j} & 0 & 0\\ 0 & 0 & e^{-h_j} & 0\\ 0 & 0 & 0 & e^{h_j} \end{bmatrix},
~~~~
\text{where} 
~~~~
    \mathcal{H} 
    = h_j \cdot\mathcal{Z}_1 \otimes\mathcal{Z}_2
    = \begin{bmatrix} h_j & 0\\ 0 & -h_j \end{bmatrix} \otimes \begin{bmatrix} 1 & 0\\ 0 & -1 \end{bmatrix}
    = \begin{bmatrix} h_j & 0 & 0 & 0\\ 0 & -h_j & 0 & 0\\ 0 & 0 & -h_j & 0\\ 0 & 0 & 0 & h_j \end{bmatrix}.
\ee
\end{proof}

\begin{lemma}
\label{addQubitLemma}
    Let $\mathcal{U}_n = e^{h_j \cdot \mathcal{H}_1}$ act on $n$ qubits, and $\mathcal{C}x$ denote a CNOT gate with the target on qubit $n$ and the control on a new qubit labelled $n+1$ then
    \be
    \label{RecursiveRelation}
        \mathcal{U}_{n+1} = \mathcal{C}x  \mathcal{U}_{n} \mathcal{C}x.
    \ee
\end{lemma}
\begin{proof}

Denoting $\mathcal{L}_{h_j} = \cosh{(h_j)}$ and $\mathcal{S}_{h_j} = \sinh{(h_j)},$
    $   
    \mathcal{U}_n
    =
    e^{h_j} \left( \mathcal{L}_{h_j} \mathcal{I}_1 \mathcal{I}_2 \cdots \mathcal{I}_n+ \mathcal{S}_{h_j}  \mathcal{Z}_1 \mathcal{Z}_2 \cdots \mathcal{Z}_n \right).
    $
For each of these two terms, the circuit identities in Fig.~(\ref{fig:IIdentity}) show that conjugating them with CNOTs increases $n$ by $1$.

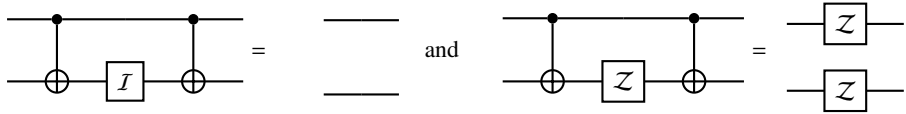
\begin{figure}[h!]
\begin{center}
\begin{quantikz}
& \ctrl{1}	& \qw & \ctrl{1} & \qw \\
& \targ{}	& \gate{\mathcal{I}} & \targ{} & \qw
\end{quantikz}
=
\begin{quantikz}
& \ghost{I}	& \qw & \qw\\
& \ghost{I}	& \qw  & \qw
\end{quantikz}
~~~\text{and}
~~~
\begin{quantikz}
& \ctrl{1}	& \qw & \ctrl{1} & \qw \\
& \targ{}	& \gate{\mathcal{Z}} & \targ{} & \qw
\end{quantikz}
=
\begin{quantikz} [row sep=0.35cm]
	& \gate{\mathcal{Z}} & \qw\\
	& \gate{\mathcal{Z}}  & \qw
\end{quantikz}
\end{center}
\caption{Circuit identities for conjugation by CNOTs.}
 \label{fig:IIdentity}
\end{figure}
\end{proof}

\subsection{Implementing $e^{\mathcal{H}_2}$}
A quantum circuit implementing $e^{\mathcal{H}_2}$ can be obtained by combining Lemmas \ref{addQubitLemma} and \ref{Lem:2to1}.
It leads to circuits of the form of Fig.~(\ref{fig:nonUnitGadgetRed}), reminiscent of Ref.~\cite{Lloyd}.

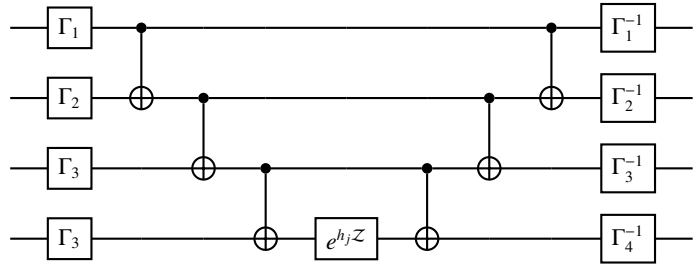
\begin{figure}[h!]
\begin{center}
\begin{quantikz}[row sep=0.3cm]
\qw 	& \gate{\Gamma_1}  & \ctrl{1}	& \qw 	& \qw 	&	\qw 					&	\qw 	& \qw 	& \ctrl{1} &  \gate{\Gamma_1^{-1}}  & \qw	\\
\qw 	& \gate{\Gamma_2}  & \targ{}	& \ctrl{1}    & \qw  	&	\qw 					&	\qw 	& \ctrl{1}	& \targ{} & \gate{\Gamma_2^{-1}}  & \qw	\\
\qw 	& \gate{\Gamma_3}  &	 \qw 		& \targ{}	& \ctrl{1} 	& \qw 				& \ctrl{1}   & \targ{} 	& \qw	&  \gate{\Gamma_3^{-1}}  &\qw		\\
\qw 	& \gate{\Gamma_3}  &	 \qw 		&  \qw 	& \targ{}	& \gate{e^{h_j\mathcal{Z}}} & \targ{}  & \qw		& \qw	& \gate{\Gamma_4^{-1}}  & \qw		
\end{quantikz}
\end{center}
\caption{Non-unitary gadget for 4 qubits.}
    \label{fig:nonUnitGadgetRed}
\end{figure}

\section{Proof of Lemma $\ref{expZtheorem}$}
\label{expZtheoremproof}
\label{ValidProof}
\begin{proof}
For $r \in \mathbb{R},$ expanding
\be
    \exp(r \FP) 
    =
    \sum^{\infty}_{j = 0} \bigg( \dfrac{\big( r \FP \big)^j}{j!} \bigg)
    =
    \sum^{\infty}_{j = 0} \bigg( \dfrac{\big( r\FP\big)^{2j}}{(2j)!} + \dfrac{\big( r\FP\big)^{2j+1}}{(2j+1)!} \bigg)
    =
    \sum^{\infty}_{j = 0} \bigg( \dfrac{r^{2j} \big( \FP^2 \big)^{j}}{(2j)!} + \dfrac{r^{2j+1} \big( \FP^2 \big)^{j} \FP }{(2j+1)!} \bigg).
\ee
As $\FP^2 = \FI$, 
\begin{align}
    \exp(r \FP) 
    =
    \sum^{\infty}_{j = 0} \bigg( \dfrac{r^{2j} \FI}{(2j)!} + \dfrac{r^{2j+1}  \FP }{(2j+1)!} \bigg)
    =
    \sum^{\infty}_{j = 0} \bigg( \dfrac{r^{2j}}{(2j)!} \bigg) \FI + \sum^{\infty}_{j = 0} \bigg( \dfrac{r^{2j+1} }{(2j+1)!}\bigg)   \FP  
    =    (\cosh r)\FI+ (\sinh r )      \FP .
\end{align}
\end{proof}

\section{Classical algorithm for partition function estimation to multiplicative error}
\label{sampBoundProof}
Our algorithm takes as inputs a $N$-qubit Hamiltonian $\mathcal{H} =  \sum^L_{j=1} h_j \FP_j$ described by $h_j, \FP_j, L,$
 a real inverse temperature $\beta,$ 
 the multiplicative sampling error $\mse,$ 
 the upper bound on the probability  $\delta$ of obtaining an estimate beyond this precision,
 the number of Trotter steps $\nu_{\mathrm{B}}$ given by the RHS of Eq.~\ref{eq:TrotNum} achieving a designated multiplicative Trotter error in the estimate of the partition function, 
 as well as its maximum value $ Z_{\mathrm{max}}.$
It is identical to Algorithm 3 in Ref.~\cite{chowdhury2019computing}, except for the exponential factors in the last line.

\begin{algorithm}[H]
\SetAlgoLined
\KwInput{$h_j, \FP_j, L, N, \beta, \nu_{\mathrm{B}}, Z_{\mathrm{max}}, \delta, \mse $}

Set ApproxNum = 0

ZR = $Z_{\mathrm{max}}$ 

RequiredTrotterSteps = $\nu_{\mathrm{B}}$

ZRPrime = 0

\While{$\mathrm{ZR} \geq \mathrm{ZRPrime}$}{
    \hspace{2em} ApproxNum++
    
    \hspace{2em} ZR = $\dfrac{Z_{\mathrm{max}}}{2^{\mathrm{ApproxNum}}}$
    
    \hspace{2em} absError = $\dfrac{\mse}{2} \times \mathrm{ZR}$
    
    \hspace{2em} deltaPrime =  $\dfrac{6\delta}{\pi^2 \times \mathrm{ApproxNum}^2}$
    
    \hspace{2em} Fraction = AES($h_j, \FP_j, \beta, \nu_{\mathrm{B}}, L, N, \delta \ase$)					 \tcp*{Algorithm \ref{AddSample}}
    						
    \hspace{2em} ZRPrime = Fraction  $\times ~2^N  \exp(\beta \Omega)$			
}
\KwOutput{ZRPrime}
 \caption{Algorithm for partition function estimation to multiplicative (relative) error }
\label{algorithmFormal}
\end{algorithm}

\begin{algorithm}[H]
\label{AddSample}
\SetAlgoLined
\KwInput{$h_j, \FP_j, \beta, \nu_{\mathrm{B}}, L, N, \delta, \ase$}

Set SAMPLES = $\emptyset$\\

numSamples = $ \left \lceil -2\ln(\delta)/\epsilon_{\textit{abs}}^2 \right \rceil $\\
$k = 1$\\
\For{$k \leq \mathrm{numSamples}$}
{
\hspace{2em} SAMPLE =  SSA($h_j, \FP_j, \beta, \nu_{\mathrm{B}}, L, N$) \tcp*{Algorithm \ref{SingleSample}}
\hspace{2em}Add SAMPLE to SAMPLES\\
\hspace{2em} $k += 1$\\
}
\KwOutput{Average of SAMPLES}
\caption{Additive error samples (AES) \label{AdditiveErrorSamples}}
\end{algorithm}

\begin{algorithm}[H]
\label{SingleSample}
\KwInput{$h_j, \FP_j, \beta, \nu_{\mathrm{B}}, L, N$}

 Set CIRCUIT $= \FI.$

 \While{($k \leq \nu_{\mathrm{B}}$)}{
   \hspace{2em}\For{($j \leq L$)}{
       \hspace{4em}Set RAND to a uniformly random number in $[0, 1]$\\
         \hspace{4em}\If{$ \left( \mathrm{RAND} \geq \dfrac{ \sinh |c_j |}{e^{\vert c_j \vert}} \right)$}{
            \hspace{6em}CIRCUIT = CIRCUIT  $\times ~ \Sigma(c_j) \FP_j $}
 }
 }
DQC1 = Circuit in Fig.~\ref{fig:TraceEstimation} applied with $U=$ CIRCUIT\\
SAMPLE = Outcome of classically evaluating DQC1 on input uniformly selected from all $N$-bit binary strings.\\
\KwOutput{SAMPLE}
\caption{Single Sample Algorithm (SSA) \label{SingleSampleAlgorithm}}
\end{algorithm}

\section{Proof of Theorem \ref{TraceErrorTheorem}}
\label{errorBoundProof}
This Appendix uses methods from Ref.~\cite{childs2019theory} to prove Theorem \ref{TraceErrorTheorem}.
Other methods~\cite{clinton2020Hamiltonian,PhysRevA.102.010601} may also be used.

\subsection{Preparatory Lemmas}
\begin{lemma}[Inverse Product]
\label{InverseProduct}
For a sequence of operators $\{A_1,\cdots, A_L\}$ and $j,k \in \{1,\cdots,L\},$
\be
        \prod^L_{j = k} \left( e^{\tau A_j} \right) \prod^1_{j = L} \left( e^{-\tau A_j} \right) = \prod^1_{j = k-1}\left( e^{-\tau A_j} \right).
 \ee
 \end{lemma}

\begin{proof}
Noting that
    \begin{align}
        \prod^L_{j = k} \left( e^{\tau A_j} \right) \prod^1_{j = L} \left( e^{-\tau A_j} \right) 
        = \prod^{L-1}_{j = k} \left( e^{\tau A_j} \right) ~ e^{\tau A_{L}} e^{-\tau A_{L}}  \prod^1_{j = L-1} \left( e^{-\tau A_j} \right)
        = \prod^L_{j = k} \left( e^{\tau A_j} \right)   \prod^1_{j = L-1} \left( e^{-\tau A_j} \right),
    \end{align}
the result follows by repeatedly redefining $L$, and iterating.
\end{proof}

\begin{lemma}[Lemma A.3 of Ref.~\cite{childs2019theory}]
\label{Fundamental}
Let $\U (t)$ be an invertible and continuously differentiable operator-valued function for $t \in \mathbb{R}.$
Then there exists $ \G (t)$ such that
\be
    \U(t) = \exp{\bigg\{ \int^{t}_0 \bigg( \G(\tau) \bigg)d \tau \bigg\}}\U(0),
    ~~~\text{and}~~~
    \G(\tau) = \dfrac{d}{d \tau}\bigg( \U(\tau) \bigg)~\U^{-1}(\tau)
\ee
\end{lemma}

\begin{proof} 
    Using the chain rule on $\U(t)$ which is invertible and continuously differentiable, let 
    \be
        \G(\tau)
        =
        \dfrac{d}{d\tau} \bigg(  \ln{(\U(\tau))} \bigg) = \dfrac{d}{d \tau}\bigg( \U(\tau) \bigg)\U^{-1}(\tau) .
  \ee
  Then
    \be 
        \int^{t}_0 \bigg( \G(\tau) \bigg)~d \tau
        = \int^{t}_0 \bigg( \dfrac{d}{d\tau} \bigg(  \ln{(\U(\tau))} \bigg) \bigg)~d \tau
        = \ln{(\U(t))} - \ln{(\U(0))}
    \ee
    whereby
    $
        \exp{\bigg\{ \int^{t}_0 \bigg( \G(\tau) \bigg)d \tau \bigg\}}~\U(0) = \exp{\bigg\{ \ln{(\U(t))} - \ln{(\U(0))} \bigg\}} ~\U(0).
    $
    As $\U(t)$ is invertible, $ \ln{(\U(t))} - \ln{(\U(0))} = \ln{(\U(t)) ~ \U^{-1}(0))}$ and
    $
        \exp\left\{ \int^{t}_0 \bigg( \G(\tau) \bigg)d \tau \right\}~\U(0) = \exp\left\{ \ln{(\U(t)~\U^{-1}(0))} \right\}\U(0) = \U(t).
    $
\end{proof}

\begin{lemma}[Lemma A.2 of Ref.~\cite{childs2019theory}]
\label{sums}
 If $\A(\beta)$ and $\B(\beta)$ are continuous operators, then
    \be
        exp \left\{ \int^{\beta}_0 \left( \A(\tau) + \B(\tau) \right) d \tau \right\}
        = \exp\left\{ \int^{\beta}_0 \A(\tau) ~ d \tau \right\} 
        \cdot \exp \left\{ \int^{\beta}_0 \left( \exp\left\{ - \int^{\tau_1}_0  \A(\tau_2)  ~d \tau_2 \right\} 
        \cdot \B(\tau_1) \cdot \exp\left\{ \int^{\tau_1}_0 \left( \A(\tau_2) \right) d \tau_2 \right\} \right) d \tau_1 \right\}
    \ee
\end{lemma}

\subsection{From operator Trotter error to scalar multiplicative (relative) error }

\begin{lemma}
\label{singToEig}
    Let $\mu_j( \cdot )$, and $\sigma_j( \cdot )$ be the $j$'th smallest singular value and eigenvalue, respectively, of their arguments.
    Then, for a Hamiltonian $\mathcal{H}, \beta \in \mathbb{R}$ and $\nu \in \mathbb{Z}^+,$
    \be 
        \left[ \mu_j \left( \exp \left(-\frac{\beta}{\nu} \mathcal{H}\right) \right) \right]^{\nu}
        =
        \sigma_j \left( \exp (-\beta \mathcal{H} ) \right).
    \ee
\end{lemma}
    
\begin{proof}
As $\exp \left(-\frac{\beta}{\nu} \mathcal{H} \right)$ is positive semi-definite,
$
    \mu_j \left(\exp \left(-\frac{\beta}{\nu} \mathcal{H} \right) \right) = \sigma_j \left(\exp \left(-\frac{\beta}{\nu} \mathcal{H} \right) \right).
$
Furthermore, as 
$
\sigma_j \left(\exp \left(-\frac{\beta}{\nu} \mathcal{H} \right) \right) = \exp \left(-\frac{\beta}{\nu} \sigma_j \left( \mathcal{H} \right) \right),
$ 
the result follows from raising both sides to the power $\nu.$
\end{proof}

Recall from Appendix~\ref{DQC1Error} that $\oTR$ denotes one Trotter step for the imaginary time evolution.
As this Appendix deals exclusively with real temperatures which correspond to imaginary time evolutions, we suppress the subscript $R$ 
and denote $\oT \equiv \oTR$ for brevity. Note that the number of Trotter steps $\nu$ is implicit in $\oT$ (see Eq.~(\ref{eq:trotapp})).
We denote the operator error $\w_{\nu}$ in one Trotter step as 
\be
    \exp \left(-\frac{\beta}{\nu} \mathcal{H} \right) \w_{\nu} = \oT \equiv \prod^L_{j=1}  \exp \left( - \frac{\beta}{\nu} h_j \FP_j \right) .
\ee

\begin{lemma} 
\label{ProperTraceBound}
  The multiplicative (relative) Trotter error $\mte$ in Eq.~(\ref{eq:multerr}) in approximating  $Z = \tr \left( \exp(- \beta \mathcal{H}) \right) $ 
by $\zt = \tr \left([\oT]^{\nu} \right)$ is bounded by
$
  \mte \geq \SN{\w_{\nu}}^{\nu} -1.
$
\end{lemma}

\begin{proof}
Using Lemma $\ref{singToEig}$, the sub-multiplicity of the spectral norm, and that the spectral norm is greater than or equal to all singular values
\be
    \zt
    =  \tr \left([\oT]^{\nu} \right)
    = \tr \left(\left[ \exp \left(-\frac{\beta}{\nu} \mathcal{H} \right) \w_{\nu}\right]^{\nu} \right)
    \leq   \sum^{2^N}_{j = 1}   \sigma_j \left( \exp (-\beta \mathcal{H} ) \right) \SN{\w_{\nu}}^{\nu}
    = Z \SN{\w_{\nu}}^{\nu},
\ee
where the Hamiltonian $\mathcal{H}$ acts on $N$ qubits. 
Thus, $| Z - \zt  | \leq \left(\SN{\w_{\nu}}^{\nu} -1 \right)Z.$
The result follows by setting   $\mte \geq \SN{\w_{\nu}}^{\nu} -1$ in Eq.~(\ref{eq:multerr}).
\end{proof}

\subsection{Proof of Theorem $\ref{TraceErrorTheorem}$}
The following proof follows the strategy of Ref.~\cite{childs2019theory} to exploit the non-commutativity amongst the terms of Hamiltonian.
However, for simplicity, it is specialised to the first-order Trotterization.
It uses Lemma~\ref{Fundamental} to set up the use of Lemma~\ref{sums}.
This gives us an expression of an operator essentially applying the error $\mathcal{W}(\beta).$
This then leads to a bound of the spectral norm of the error operator in a single Trotter step $\mathcal{W}(\beta/\nu) = \w_{\nu}.$

\begin{proof}
Denote $A_j \equiv h_j \FP_j,$ and
\be
    \Theta(\beta) = \prod^{L}_{j = 1} \left( e^{-\beta A_j} \right).
\ee
Note that 
$
    \Theta(\beta/\nu) = \oT  = \prod^L_{j=1}  \exp \left( - \frac{\beta}{\nu} h_j \FP_j \right) .
$
By differentiation,
\begin{align}
    \dfrac{d \Theta(\beta)}{d\beta} 
    &= -A_1 \Theta(\beta) - \sum^L_{k=2} \left[ \prod^{k-1}_{j = 1} \left( e^{-\beta A_j} \right) \cdot A_k \cdot \prod^{L}_{j = k} \left( e^{-\beta A_j} \right) \right]\\
    &= \left\{ - A_1 - \sum^L_{k=2} \left[ \prod^{k-1}_{j = 1} \left( e^{-\beta A_j} \right) \cdot A_k \cdot \prod^{L}_{j = k} \left( e^{-\beta A_j} \right)\right] \prod^1_{j = L} \left( e^{\beta A_j} \right) \right\} \Theta(\beta) \\
    &= \left\{ - A_1 - \sum^L_{k=2} \left[ \prod^{k-1}_{j = 1} \left( e^{-\beta A_j} \right) \cdot A_k \cdot \prod^1_{j = k-1}\left( e^{\beta A_j} \right) \right] \right\} \Theta(\beta),
\end{align}
where we have used Lemma~\ref{InverseProduct} in the last step.
Denoting
\be
    \label{eq:Sk}
    \mathcal{F}(\beta) 
    \equiv - A_1 -  \sum^L_{k=2}\s_k (\beta) ~~~\text{with}~~~
    \s_k(\beta) = \prod^{k-1}_{j = 1} \left( e^{-\beta A_j} \right) \cdot A_k \cdot \prod^1_{j = k-1}\left( e^{\beta A_j} \right),
\ee
whereby
\begin{align}
    \dfrac{d \Theta(\beta)}{d\beta} 
    &= \mathcal{F}(\beta) \Theta(\beta).
\end{align}
Lemma \ref{Fundamental} gives 
\begin{align}
    \Theta(\beta) 
    &= \exp \left\{ \int^{\beta}_0 \left( \dfrac{d \oT(\tau)}{d\tau} \oT^{-1}(\tau) \right) d \tau \right\}
    = \exp \left\{ \int^{\beta}_0\mathcal{F}(\tau) ~ d \tau \right\}
    = \exp \left\{ \int^{\beta}_0 \left( -\mathcal{H} + \left[ \mathcal{F}(\tau) +\mathcal{H}\right] \right) ~d \tau \right\}.
\end{align} 
Lemma \ref{sums} gives
\begin{align}
    \Theta(\beta) 
    &=
    \exp\left\{ \int^{\beta}_0 ( -\mathcal{H}) ~d \tau \right\} 
    \exp\left\{ \int^{\beta}_0 \left( \exp \left\{ \int^{\tau_1}_0 \mathcal{H}  ~ d \tau_2 \right\} 
	\left[ \mathcal{F}(\tau_1) + \mathcal{H} \right] 
	\exp \left\{ \int^{\tau_1}_0 \left( -\mathcal{H} \right) ~ d \tau_2 \right\} \right) d \tau_1 \right\}\\
    &=
    \exp \left\{ -\beta \mathcal{H} \right)
    \exp\left\{ \int^{\beta}_0 \left( e^{\mathcal{H}\tau_1}  \left[ \mathcal{F}(\tau_1) + \mathcal{H} \right] e^{- \mathcal{H}\tau_1} \right) ~d \tau_1 \right\} \\
    &\equiv
    \exp \left\{ -\beta \mathcal{H} \right)\mathcal{W}(\beta),
\end{align}
where $\mathcal{W},$ denotes the error operator when $\exp \left\{ -\beta \mathcal{H} \right)$ is approximately implemented via $ \Theta(\beta).$
Then using Eq.~(\ref{eq:Sk})
\begin{align}
    \mathcal{W} (\beta)
    &=
    \exp \left \{ \int^{\beta}_0 \left( e^{\mathcal{H}\tau_1}  \left[ \mathcal{F}(\tau_1) + \mathcal{H} \right]  e^{ -\mathcal{H}\tau_1} \right) ~ d \tau_1 \right\}
    =
    \exp\left \{ \int^{\beta}_0 \left( e^{\mathcal{H}\tau_1}  \mathcal{F}(\tau_1)  e^{ -\mathcal{H}\tau_1}  + \mathcal{H} \right) ~ d \tau_1 \right\}\\
    \label{EOther}
    &=
    \exp\left \{ \int^{\beta}_0 \left( e^{\mathcal{H}\tau_1} \left[ - A_1 - \sum^L_{k=2} \s_k(\tau_1) \right]  e^{ -\mathcal{H}\tau_1}  + \mathcal{H} \right) ~ d \tau_1 \right\}.
\end{align}

Let $z$ be the greatest index in $ \mathcal{\zeta}_k$ less than $k.$
Then 
\begin{align}
    \label{psiDef}
    \s_k (\tau_1)
    &=
    \prod^{k-1}_{j = 1} \left( e^{-\tau_1 A_j} \right) \cdot A_k \cdot \prod^1_{j = k-1}\left( e^{\tau_1 A_j} \right)
    =
    \prod^{z-1}_{j = 1} \left( e^{-\tau_1 A_j} \right) 
    \cdot e^{-\tau_1 A_{z}} A_k e^{\tau_1 A_{z}}
    \cdot \prod^1_{j = z - 1}\left( e^{\tau_1 A_j} \right)\\
    \label{psiDef}
    &=
    \prod^{z - 1}_{j = 1} \left( e^{-\tau_1 A_j} \right) \cdot A_k \cdot \prod^1_{j = z - 1}\left( e^{\tau_1 A_j} \right) \\
    &+
    \prod^{z - 1}_{j = 1} \left( e^{-\tau_1 A_j} \right) \cdot \left[ A_{z}, A_k \right] \tau_1 \cdot \prod^1_{j = z - 1} \left( e^{\tau_1 A_j} \right)
    +
    \prod^{z-1}_{j = 1} \left( e^{-\tau_1 A_j} \right) \cdot \int^{\tau_1}_0 \int^{\tau_2}_0 \left( e^{-\tau_3 A_{z}} \left[ A_{z}, \left[ A_{z}, A_k \right] \right] e^{\tau_3 A_{z}}\right) ~ d \tau_3 d \tau_2 \cdot \prod^1_{j = z - 1}\left( e^{\tau_1 A_j} \right),
\end{align}
where this last line follows from the identity
\be
    e^{t A} B e^{- t A} = B + [A,B] t + \int_0^t d t_2 \int_0^{t_2} d t_3 e^{t_3 A} [A, [A,B]] e^{-t_3 A} .
\ee
Note that the first term, in Eq.~(\ref{psiDef}), is devoid of $A_{z}.$ 
Repeating the above on the first term, 
\begin{align}
    \label{ReturntoPsi}
    \s_k (\tau_1) = A_k + \D_k (\tau_1),
\end{align}
where
\be
    \D_k (\tau_1) =  \sum_{\substack{z \leq k \\ z \in \mathcal{\zeta}_k}}
    \left[ \prod^{z - 1}_{j = 1} \left( e^{-\tau_1 A_j} \right) \left[ A_{z}, A_k \right]~ \tau_1  \prod^1_{j = z - 1} \left( e^{\tau_1 A_j} \right)
    + \prod^{z-1}_{j = 1} \left( e^{-\tau_1 A_j} \right)  \int^{\tau_1}_0 \int^{\tau_2}_0 
    \left( e^{-\tau_3 A_{z}} \left[ A_{z}, \left[ A_{z}, A_k \right] \right] e^{\tau_3 A_{z}} \right) d \tau_3 d \tau_2 
     \prod^1_{j = z - 1}\left( e^{\tau_1 A_j} \right) \right].
\ee
Thus, the quantity in the square brackets in Eq.~(\ref{EOther}) equals $\mathcal{H} + \sum_{k=2}^L \D_k(\tau_1),$ whereby
\be
    \mathcal{W} (\beta) = \exp\left \{ - \int^{\beta}_0 \left( e^{\mathcal{H}\tau_1} \left[  \sum^L_{k=2} \D_k(\tau_1) \right]  e^{ -\mathcal{H}\tau_1}   \right) ~ d \tau_1 \right\}.
\ee
Thus,
\be
    \SN{ \mathcal{W} (\beta)} 
    \leq
    \exp\left \{ \int^{\beta}_0 \left( e^{\SN{\mathcal{H} }\tau_1} \left[  \sum^L_{k=2} \SN{\D_k(\tau_1) }\right]  
	e^{ \SN{\mathcal{H}} \tau_1}   \right) ~ d \tau_1 \right\}.
\ee
As the commutator of Paulis are Paulis, and their spectral norm is unity,
$
    \SN{[A_{z}, A_k]} = 2|h_{z}| |h_k| 
$
and
$
    \SN{\left[ A_{z}, \left[ A_{z}, A_k \right] \right] } =  4 |h_{z} |^2 |h_k|.
$
Denoting
\be
    \Omega_{k} = \sum_{j = 1}^{k} \SN{A_j}  =  \sum_{j = 1}^{k} |h_j|,~~~\text{and}~~~\Omega \equiv \Omega_L,
\ee
\be
    \SN{\D_k(\tau_1) } \leq  2  |h_{z}| |h_k|  \sum_{\substack{z \leq k \\ z \in \mathcal{\zeta}_k}} 
    \left[
		e^{ 2 \tau_1\Omega_{z - 1} } \tau_1 
		+ 2 |h_{z}| e^{ 2 \tau_1\Omega_{z-1} }
		\int^{\tau_1}_0 \int^{\tau_2}_0 e^{2 |h_{z} | \tau_3 } d \tau_3 d \tau_2 
	\right]
	=
	|h_k| \sum_{\substack{z \leq k \\ z \in \mathcal{\zeta}_k}} 
	\left[    
	    e^{ 2 \tau_1\Omega_{z}} - e^{ 2 \tau_1\Omega_{z-1} }
	\right]	 .     
\ee
Denoting $\omega_z = 2(\Omega + \Omega_z),$
\be
    \SN{ \mathcal{W} (\beta)} 
    \leq 
    \exp\left \{
        \sum^L_{k=2} |h_k|  \sum_{\substack{z \leq k \\ z \in \mathcal{\zeta}_k}} 
        \left[    
	        \frac{e^{ \beta \omega_z}-1}{\omega_z} - \frac{e^{\beta \omega_{z-1}}-1}{\omega_{z-1}} 
	    \right]
    \right\}.
\ee
Setting $w = 2\beta \Omega/\nu,$  the spectral norm of the Trotter error in $\nu$ Trotter steps of size $\beta/\nu$ each is
\be
    \SN{ \mathcal{W}_{\nu} }^{\nu} 
    \leq 
    \exp\left \{ 
        \beta \mathfrak{h}
 		\left[    
		    \frac{e^{ 2w} - 1}{2w} - \frac{e^{w} -1}{w}
		\right] 
	\right\}
    \approx	
    \exp\left \{
        \beta \mathfrak{h} \frac{w}{2} 
    \right\},
\ee
where $\mathfrak{h}= \sum^L_{k=2} |h_k| N_k,$ with $N_k = |\{z \leq k \text{~such that~} z \in \mathcal{\zeta}_k \} |$ 
accounting for the mutual non-commutativity amongst the terms of $\mathcal{H},$
The last approximation neglects higher order terms in $w$ because
for a fixed Hamiltonian ($\Omega$) and temperature ($\beta$), $w \rightarrow 0$ as $\nu \rightarrow \infty.$
Finally, using Lemma \ref{ProperTraceBound}, 
\be
    \nu \geq \frac{\beta^2 \Omega \mathfrak{h} } {\ln(1+\mte)}.
\ee
\end{proof}

\section{Hardness of Pauli decomposition}
\label{HardnessProof}
\label{Decomposition}
We begin with a formal statement of the problem.

\begin{definition}[Obtaining a H-decomposition]
Given an $N$-qubit Hermitian operator $H $ as a set of $\poly(N)$ positive coefficients $c_i$ and unitary operators $U_i$, the latter described by $\poly(N)$-sized quantum circuits each such that $H = \sum_i c_i U_i,$ a $N$-qubit Pauli operator $\sigma$,  $ \delta \geq 0$, $\delta = \poly(n)$, and a bit Re or Im,
the goal is to output either $\hat{r}$ or $\hat{t}$, depending on whether the bit is Re or Im, such that
\begin{eqnarray}
| \hat{r} - \text{Re} (\tr( \sigma H)) | \leq \delta, \\
| \hat{t} - \text{Im} (\tr ( \sigma H)) | \leq \delta.
\end{eqnarray} 
\end{definition}

We also define a problem of known hardness - that of estimating up to polynomial additive error the real or imaginary part of the coefficient of a unitary in its Pauli expansion.

\begin{definition}[Obtaining a U-decomposition]
Given an $N$-qubit unitary operator $U$ as a $\poly(N)$-sized quantum circuit, an $N$-qubit Pauli operator $\sigma$, $ \delta\geq0, \delta = Poly(n)$, and a bit Re or Im, the goal is to output either $\hat{r}$ or $\hat{t}$, depending on whether the bit is Re or Im, such that
\begin{eqnarray}
    | \hat{r} - \text{Re} (\tr ( \sigma U)) | \leq \delta\\
    | \hat{t} - \text{Im} (\tr ( \sigma U)) | \leq \delta.
\end{eqnarray} 
\end{definition}

\begin{theorem}[\cite{Knill_1998}]
Obtaining a U-decomposition is DQC1-hard.
 \label{knill}
\end{theorem}

We prove that obtaining a U-decomposition can be polynomially transformed into obtaining a H-decomposition, thus the hardness of the latter.

\begin{theorem}
Obtaining a U-decomposition $\leq_p$ Obtaining a H-decomposition. Therefore, obtaining a H-decomposition is DQC1-hard.
\end{theorem}

\begin{proof}
It suffices to provide a polynomial-time algorithm for transforming the inputs to the problem of obtaining a U-decomposition into the inputs of obtaining a H-decomposition problem, such that the transformed problem has same output as the original.

This algorithm does the following: It takes as input the circuit description of $U$. It generates the description of $U^{\dagger}$ by reversing the circuit and conjugating all the gates.
In the case we are asked to calculate the real part of the coefficient corresponding to $\sigma$, the algorithm generates the output 
\be
    \left\{\left\{\frac{1}{2}, U\right\}, \left\{\frac{1}{2}, U^{\dagger} \right\} \right\}, \sigma, \delta, \text{Re}.
\ee
This is a valid input to the problem of obtaining a H-decomposition as $\dfrac{1}{2} (U+U^{\dagger})$ is Hermitian and the descriptions of the unitary operators are of polynomial size.
We also have
\be
    \text{Re}\left[ \tr \left(\sigma \frac{U+U^{\dagger}}{2} \right) \right]
    = \frac{\text{Re}[ \tr(\sigma U)] + \text{Re} [\tr(( U \sigma)^{\dagger})] }{2}  
    = \frac{\text{Re}[ \tr(\sigma U)] + \text{Re} [\tr( \sigma U)^*] }{2}
    = \text{Re}[\tr( \sigma U)].
\ee
Thus, the output of the transformed problem will be the same estimate with the same precision as in the original problem.

Similarly, if we are asked to calculate the imaginary part of the coefficient that corresponds to Pauli $\sigma$, the algorithm generates the output 
\be
    \left\{\left\{-\frac{i}{2}, U\right\}, \left\{\frac{i}{2}, U^{\dagger} \right\} \right\}, \sigma, \delta, \text{A bit specifying the real or imaginary part} .
\ee
Again, this is a valid input to the problem of obtaining a H-decomposition because $\dfrac{i}{2} (U^{\dagger}-U)$ is Hermitian and the descriptions of the unitary operators are of polynomial size.
We  also have
\be
    \text{Re}\left[ \tr \left(i \sigma \frac{U^{\dagger} - U}{2} \right) \right]
    = \frac{\text{Re}[-i \tr(\sigma U)] + \text{Re} [\tr(( -i U \sigma)^{\dagger})] }{2}  
    = \frac{\text{Re}[ \tr(-i \sigma U)] + \text{Re} [\tr( -i \sigma U)^*] }{2}
    = \text{Im}[\tr( \sigma U)].
\ee
and the same reasoning as before applies for the output.
The hardness follows directly from Theorem~\ref{knill}.
\end{proof}

\end{widetext}
\end{document}